# Tailoring the electronic properties of SrRuO$_3$ films in SrRuO$_3$/LaAlO$_3$ superlattices

Z. Q. Liu,[1,2] M. Yang,[2] W. M. Lü,[1,3] Z. Huang,[1] X. Wang,[1,2] B. M. Zhang,[1,4] C. J. Li,[1,5] K. Gopinadhan,[1,3] S. W. Zeng,[1,2] A. Annadi,[1,2] Y. P. Feng,[1] T. Venkatesan,[1,2,3] and Ariando[1,2*]

[1]*NUSNNI-Nanocore, National University of Singapore, Singapore 117411*

[2]*Department of Physics, National University of Singapore, Singapore 117542*

[3]*Department of Electrical and Computer Engineering, National University of Singapore, Singapore 117542*

[4]*Departement of Material Science Engineering, National University of Singapore, Singapore 117576*

[5]*National University of Singapore (NUS) Graduate School for Integrative Sciences and Engineering, 28 Medical Drive, Singapore 117456.*

**Abstract:** The electronic properties of SrRuO$_3$/LaAlO$_3$ (SRO/LAO) superlattices with different interlayer thicknesses of SRO layers were studied. As the thickness of SRO layers is reduced, the superlattices exhibit a metal-insulator transition implying transformation into a more localized state from its original bulk metallic state. The strain effect on the metal-insulator transition was also examined. The origin of the metal-insulator transition in ultrathin SRO film is discussed. All the superlattices, even those with SRO layers as thin as 2 unit cells, are ferromagnetic at low temperatures. Moreover, we demonstrate field effect devices based on such multilayer superlattice structures.

*To whom correspondence should be addressed. E-mail: ariando@nus.edu.sg

Metal oxides are essential to oxide electronic devices typically as electrodes and templates for integration of other oxide materials, such as in the resistive switching[1] and various tunnel junction devices.[2–4] SrRuO$_3$ (SRO) is a conductive ferromagnetic oxide,[5] which has been extensively utilized as ferromagnetic electrodes in magnetic tunnel junctions[6] and normal metal electrodes in resistive switching devices.[7] Moreover, field effect devices based on ultrathin SRO films has been demonstrated by Ahn *et al.*.[8] Generally, technological scaling of electronic devices stresses the need of using metal oxides in terms of ultra thin films. However, an important issue on using ultrathin films of metal oxides is the substantially increased resistivity, which is clearly manifested in SRO thin films.[8–10] For example, Ahn *et al.*[8] have found that room temperature resistivity of a 2 nm SRO film grown on a SrTiO$_3$ (STO) substrate reaches 20 times the bulk value; Toyota *et al.* observed a metal-insulator transition in SRO films occurring at a film thickness of 4 or 5 unit cells (uc); Later, Xia *et al.* observed a critical thickness of 4 uc, below which SRO films become insulating while the ferromagnetic character measured via magneto-optic Kerr measurements disappears.

Recently, a highly confined spin-polarized two-dimensional electron gas has been predicted in SRO-based superlattices[11] and strong antiferromagnetic interlayer exchange coupling has been observed in SRO/La$_{0.7}$Sr$_{0.3}$MnO$_3$ superlattices with ultrathin SRO layers.[12] Furthermore, Boris *et al.*[13] has demonstrated that electronic phases of LaNiO$_3$ (LNO) can be substantially tuned by varying the thickness of LNO layers in LaNiO$_3$/LaAlO$_3$ superlattices, which opens up an avenue to modulate electronic properties of strongly correlated electron systems in superlattices. The strong electron-electron correlation in SRO has been investigated by thermal and electrical transport measurements.[14–16] In addition, high quality superlattices consisting of multilayer SRO ultrathin films prevent a surface depletion typically occurring at the interface between a metallic film and air.[17] Also, a superlattice structure with multilayer SRO ultrathin films gives rise to an enhanced magnetic signal compared to that of an ultrathin SRO film with the same thickness, which can facilitate the determination of magnetic properties of an ultrathin SRO film in moment measurements. Therefore, investigating the electronic properties of ultrathin SRO films in multilayer structures can provide experimental basis for potential new functionalities and devices. In this work, we report the evolution of the electronic properties of SRO/LAO superlattices as a function of SRO layer thickness and demonstrate a field effect modulation in these SRO/LAO superlattices.

Bulk SRO is an itinerate ferromagnet with $T_c \sim 160$ K and orthorhombic with lattice parameters of $a = 5.55$ Å, $b = 5.56$ Å and $c = 7.86$ Å.[18] Its pseudocubic structure has a lattice constant of 3.93 Å[19] and therefore SRO films can be epitaxially grown on common perovskite substrates (e.g., STO or LAO, both are nonmagnetic insulators with a large band gap of 3.2 and 5.6 eV, respectively).[18,20] A series of superlattices with alternating SRO and LAO thin layers were fabricated, *i.e.*, [SRO$_{2uc}$LAO$_{2uc}$]$_{60}$, [SRO$_{3uc}$LAO$_{3uc}$]$_{40}$, [SRO$_{4uc}$LAO$_{4uc}$]$_{30}$, [SRO$_{7uc}$LAO$_{7uc}$]$_{20}$, and [SRO$_{10uc}$LAO$_{10uc}$]$_{12}$ on small-miscut TiO$_2$-terminated STO substrates by pulsed laser deposition, which are in sequence defined as SL2, SL3, SL4, SL7 and SL10, respectively. The superlattices with different SRO and LAO thicknesses were deposited at 750 °C under an oxygen partial pressure of 200 mTorr. During deposition, the fluence of the laser energy was kept at 1.5 J/cm$^2$ and the laser repetition rate was 5 Hz. The deposition rates of SRO and LAO films were carefully calibrated by performing x-ray reflectivity measurements for thick single-layer SRO and LAO films grown on STO substrates.

To avoid the effect of the two dimensional electron gas[21] at the interface between LAO and STO on electrical properties of superlattices, SRO was always the first layer deposited on the STO substrate. Both the early study by Gan *et al.*[22] on SRO films grown on large miscut (001)-oriented STO substrates (~2°) and the recently transmission electron microscopy study by Ziese *et al.*[23] on SRO films grown on small-miscut (001)-oriented STO substrates (~0.1°) show that SRO films are (110)-oriented, and the [001] and [1-10] directions of SRO films are in plane and aligned with the crystalline axes of STO. Figure 1(a) shows an atomic force microscopy (AFM) image of a (001)-oriented TiO$_2$-terminated STO substrate with a small miscut angle of ~0.1°. Such STO substrates with an average step width of 300-400 nm are similar to what Ziese *et al*. used in their study.[23]

SRO/LAO superlattices were characterized by x-ray diffraction measurements including both $\theta$-$2\theta$ scans and reciprocal space mappings. The $\theta$-$2\theta$ scan and the reciprocal space mapping of a SL7 superlattice around the STO (002) Bragg peak are shown in Fig. 1(b) and (c), respectively. Satellite peaks up to the third order can be clearly seen, suggesting a good periodicity of the superlattice. The detailed analysis of the x-ray diffraction data is provided in Ref. 24.

The electrical properties of the superlattices were characterized by typical four-probe linear resistance measurements using a Quantum Design physical property measurement system machine. Ohmic contacts onto the 5×5 mm$^2$ samples were formed using Al wedge bonding directly connected to the supperlatices. Temperature dependence of the sheet resistance ($R_s$-$T$) of different superlattice samples as well as a 50 nm thick single-layer SRO film was summarized in supplementary material.[24] It was found that the room temperature sheet resistance increases with decreasing thickness of SRO layers. And the metallic behaviour of bulk SRO finally transforms into a completely insulating state as SRO layer thickness is reduced to 2 uc in each period. Specifically, low temperature resistance upturns at 30 K and 50 K are present for SL10 and SL7 samples, respectively; and a resistance upturn at ~185 K exists for both SL4 and SL3 structures.

For a detailed analysis on the evolution of electrical properties of SRO/LAO superlattices with SRO layer thickness, $R_s$-$T$ curves of the 50 nm SRO single-layer film, SL7, SL4 and SL2 superlattices under both zero and 9 T (out-of-plane) magnetic fields are plotted on different scales in Fig. 2(a)-(d), respectively. Room temperature sheet resistance of the 50 nm SRO film is ~56 $\Omega$ and thus corresponds to a room temperature resistivity of 280 $\mu\Omega$ cm, comparable to the value in previous studies.[25] A resistivity kink at ~160 K corresponding to the Curie temperature is visible due to strong magnetic scattering, which can be clearly seen in the resistance derivative curve as shown in the inset of Fig. 2(a). Above 160 K, the sheet resistance is linearly dependent on temperature, for which SRO is referred to as a "bad metal".[26] The negative magnetoresistance (MR) under a 9 T magnetic field is apparent below 160 K due to the ferromagnetic phase. As SRO layer thickness becomes 7 uc in SRO/LAO superlattices, an obvious logarithmic increase of sheet resistance with decreasing temperature as well as a negative out-of-plane MR is seen in Fig. 2(b), typical of a weak localization.[27] While SRO layer thickness is reduced to 4 uc in each period, a resistance upturn occurs at ~185 K, and the sheet resistance at 2 K exceeds three times that of room temperature, which can be identified as metal-insulator transition similar to the case observed in *n*-type STO.[28] Finally, when the SRO layer in each period of a superlatticeis reduced to 2 uc, the whole structure turns into an insulating state. At low temperatures, the sheet resistance of the SL2 sample exceeds the resistance quantum $h/e^2$ = 25.8 kΩ, which means the electron mean free path becomes shorter than the Fermi wavelength so that quantum interference becomes predominant in electron transport properties, leading to a strong two-dimensional localization.[29]

To examine the strain effect on the metal-insulator transition in SRO thin films, we prepared similar SRO/LAO superlattices on various other substrates including (100) $(LaAlO_3)_{0.3} – (Sr_2AlTaO_6)_{0.7}$ (LSAT) and (110) $DyScO_3$ (DSC). The in-plane lattice constants of those substrates are 3.868 and 3.944 Å,[30] respectively. In addition, atomically flat surface of LSAT and DSC can be achieved by thermal annealing and wet etching. It was found that the strain can clearly influence the resistance and degree of localization of the superlattices. However, it does not change the trend of the localization evolution with the SRO thickness. The upturn temperatures of the resistance and the possible transport mechanisms in the SRO/LAO superlattices grown on different substrates are summarized in TABLE I. As can be seen from the table, the resistance upturn temperature becomes higher with decreasing the thickness of SRO layers for all different substrates, which indicates that the system becomes more localized. In addition, the reproducibility of the metal-insulator transition in SRO/LAO superlattices grown on different substrates also suggests that structural disorder is not a dominant effect since the degree of disorder should strongly depend on the lattice mismatch between substrate and SRO films.

Magnetic moments of the samples were measured by a Quantum Design superconducting quantum interference devices-vibrating sample magnetometer (SQUID-VSM) machine with the sensitivity of $10^{-8}$ emu. The magnetization of different superlattices as well as the thick SRO single-layer film is plotted as a function of temperature (*M-T*) in Fig. 3(a). In-plane magnetocrystalline anisotropy was examined for all the samples. It was found that only the thick SRO single-layer film and the SL10 sample exhibited in-plane magnetocrystalline anisotropy. For other superlattices with SRO layers thinner than 10 uc, the *M-T* curves measured along the two in-plane STO crystalline axes coincide. This is consistent with the result obtained by Ziese *et al.*[23] that the in-plane anisotropy between [1-10] and [001] directions of a 5 nm SRO film is much smaller than that of a 40 nm SRO film. Moreover, it was found that the [001] direction of SRO films grown on STO is close to the STO step direction and the perpendicular in-plane [1-10] axis of SRO films is close to the in-plane easy axis.[23] Our AFM and VSM measurements revealed that for both the 50 nm SRO film and the SL10 sample, the magnetic moment measured along the in-plane STO axis which is close to the step direction of a STO substrate, is smaller than that measured along the other perpendicular STO axis especially at low temperatures. This actually confirms the results obtained by Ziese *et al.*[23] In Fig. 3(a), the *M-T* curves along the [1-10] direction of SRO layers are shown for the 50 nm SRO film and the SL10 sample.

The ferromagnetic transition at ~160 K can be clearly seen in the *M-T* curve of the thick SRO film. The Curie temperature of the SL10 structure decreases to ~140 K, which is higher than the values obtained by Herranz *et al.*[20] ( ~110 K) and Toyota *et al.*[9] (~127 K) for 10 uc single-layer SRO films. The Curie temperatures of the SL7 and SL4 structures are similar (~135 K), which is higher than that of a 4 uc SRO single-layer film (~120 K) observed by Xia *et al.*.[10] The *M-T* curve of the SL2 sample presents a different trend from other curves, which could be due to the interlayer coupling of SRO layers through ultrathin LAO intermediate layers. A sharp transition temperature of ~110 K is seen. However, for ultrathin 2 uc single-layer SRO films, neither magnetometer[9] nor Kerr[10] measurements was able to detect any possible magnetic transition due to extremely weak signals.

To further examine the low temperature magnetic phase of all the samples, electrical MR measurements as well as field-dependent moment (*M-B*) measurements were performed at low temperatures. It was found that all the samples exhibited butterfly hysteresis loops in low temperature out-of-plane MR measurements, and also *M-B* hysteresis loops in moment measurements. The out-of-plane MR and in-plane *M-B* curves of the SL2 sample at 2 K are shown in Fig. 3(b) and (c), respectively. The hysteresis loops both in electrical and magnetic measurements confirm that the low temperature phase of such superlattices is ferromagnetic. In addition, the magnetization of superlattices is much smaller than that of the thick SRO film and further decreases with reducing SRO layer thickness, which could be due to the localization of electrons in superlattices with thin SRO layers. Since the ferromagnetism of SRO is itinerant, the localization of electrons leads to lack of electron wave function overlapping and thus the decrease of the ferromagnetic signals. Especially in the strongly localized SL2 sample, the saturation moment is ~0.15 $\mu_B$/Ru, which is one order of magnitude smaller than 1.6 $\mu_B$/Ru reported for bulk SRO.[31]

Furthermore, we constructed field effect devices based on the SL2 and SL7 samples by depositing Au electrodes on the backside of STO substrates as shown schematically in Fig. 4(a). STO substrates were employed as high-*k* dielectric material due to its extremely large dielectric constant at low temperatures and a gate voltage $V_G$ was applied from Au back electrodes. During measurements, the back $V_G$ was applied up to ±200 V and the gate leakage current was smaller than 5nA. Surprisingly, although the total film thickness of both the SL2 and SL7 samples was ~100 nm, the field modulation effect can be clearly seen in both devices (Fig. 4). For positive (or negative) $V_G$, $R_s$ decreases (or increases), which is in agreement with the addition (or removal) of *n*-type carriers to (or from) the transport channels. The hysteresis loops of both devices can be continuously recycled a number of times by the sweeping the field between ±200 V, which demonstrates the interplay between external carrier injection and internal localization. The $R_s$ in the SL2-based device varies as much as 2.1% when changing the $V_G$ from +200 to -200 V, which is much larger than that in the SL7-based device. That is because the characteristic width of the field modulation given by Thomas-Fermi screening length is inversely proportional to the one sixth power of carrier density. As the SL2 sample is more localized and expected to have less free carriers, this leads to a larger modulation length. Moreover, SRO layers in the SL2 structure are thinner than those in the SL7 structure, which further enhances the field modulation effect.

The evolution of electrical properties of SRO films from bad metal, to weak localization, then to MIT, and finally to strong localization as SRO layers become thinner and thinner could be a general evolution trend of the strongly correlated metallic systems when changing their dimensionality from 3D to 2D as also reported[13,32] for $LaNiO_3$. Although strain can apparently affect the degree of localization of a certain SRO/LAO superlattice, it does not change the trend of the localization evolution with SRO layer thickness. On the other hand, ferromagnetism is clearly present in SRO films as thin as 2uc and in bulk SRO it typically yields a kink in $R_s$-$T$ curves as shown in Fig. 3a due to strong ferromagnetic scattering. With the dimensionality of SRO films reduced, the ferromagnetic scattering could become dominant, thus leading to a metal-insulator transition. Therefore we conclude that the origin of the evolution in the electronic properties of SRO films could be either intrinsically due to dimensionality itself or the interplay between dimensionality and ferromagnetism.

In summary, we have studied the electronic properties of SRO/LAO superlattices. By varying the thickness of SRO layers in the superlattices, we are able to modulate both electrical and magnetic properties of SRO films in SRO/LAO superlattices. For example, the ferromagnetic metal SRO can be tuned into a ferromagnetic insulator with a much lower $T_c$ of ~110 K as SRO layers are reduced to 2 uc in SRO/LAO superlattices. Moreover, we have demonstrated field effect devices based on SRO/LAO superlattices, which reveals the possibility of realizing novel field effect devices based on multilayer structures.

We thank the National Research Foundation (NRF) Singapore under the Competitive Research Program (CRP) "Tailoring Oxide Electronics by Atomic Control" (Grant No. NRF2008NRF-CRP002-024), the National University of Singapore (NUS) for a cross-faculty grant, and FRC (ARF Grant No.R-144-000-278-112) for financial support.

**References:**

[1] Z.Q. Liu, D. P. Leusink, W.M. Lü, X. Wang, X. P. Yang, K. Gopinadhan, Y. L. Lin, A. Annadi, Y. L. Zhao, A. Roy Barman, S. Dhar, Y. P. Feng, H. B. Su, G. Xiong, T. Venkatesan, and Ariando, Phys. Rev. B **84**, 165106 (2011).

[2] Ariando, D. Darminto, H. -J. H. Smilde, V. Leca, D. H. A. Blank, H. Rogalla, and H. Hilgenkamp, Phys. Rev. Lett. **94**, 167001 (2005).

[3] L. Antognazza, K. Char, T.H. Geballe, L.L. H. King, and A. W. Sleight, Appl. Phys. Lett. **63**, 1005 (1993).

[4] M. Minohara, I. Ohkubo, H. Kumigashira, and M. Oshima, Appl. Phys. Lett. **90**, 132123 (2007).

[5] G. Koster, L. Klein, W. Siemons, G. Rijnders, J. S. Dodge, C. B. Eom, D. H. A. Blank, and M. R. Beasley, Rev. Mod. Phys. **84**, 253 (2012).

[6] K. Takahashi, A. Sawa, Y. Ishii, H. Akoh, M. Kawasaki, and Y. Tokura, Phys. Rev. B **67**, 094413 (2003).

[7] A. Sawa, Mater. Today **11**, 28 (2008).

[8] C.H. Ahn, R.H. Hammond, T.H. Geballe, M.R. Beasley, J.-M. Triscone, M. Decroux, Ø. Fischer, L. Antognazza, and K. Char, Appl. Phys. Lett. **70**, 206 (1997).

[9] D. Toyota, I. Ohkubo, H. Kumigashira, M. Oshima, T. Ohnishi, M. Lippmaa, M. Takizawa, A. Fujimori, K. Ono, M. Kawasaki, and H. Koinuma, Appl. Phys. Lett. **87**, 162508 (2005).

[10] J. Xia, W. Siemons, G. Koster, M. R. Beasley, and A. Kapitulnik, Phys. Rev. B **79**, 140407(R) (2009).

[11] M. Verissimo-Alves, P. García-Fernández, D. I. Bilc, P. Ghosez, and J. Junquera, Phys. Rev. Lett. **108**, 107003 (2012).

[12] M. Ziese, I. Vrejoiu, E. Pippel, P. Esquinazi, D. Hesse, C. Etz, J. Henk, A. Ernst, I.V. Maznichenko, W. Hergert, and I. Mertig, Phys. Rev. Lett. **104**, 167203 (2010).

[13] A.V. Boris, Y. Matiks, E. Benckiser, A. Frano, P. Popovich, V. Hinkov, P. Wochner, M. Castro-Colin, E. Detemple, V.K. Malik, C. Bernhard, T. Prokscha, A. Suter, Z. Salman, E. Morenzoni, G. Cristiani, H.-U. Habermeier, and B. Keimer, Science **332**, 937 (2011).

[14] P. Allen, H. Berger, O. Chauvet, L. Forro, T. Jarlborg, A. Junod, B. Revaz, and G. Santi, Phys. Rev. B**53**, 4393 (1996).

[15] G. Cao, S. McCall, M. Shepard, and J.E. Crow, Phys. Rev. B **56**, 321 (1997).

[16] A.P. Mackenzie, J. W. Reiner, A. W. Tyler, L. M. Galvin, S. R. Julian, M. R. Beasley, T. H. Geballe, and A. Kapitulnik, Phys. Rev. B **58**, R13318 (1998).

[17] A. Ohtomo and H.Y. Hwang, Appl. Phys. Lett.**84**, 1716 (2004).

[18] X.D. Wu, S. R. Foltyn, R.C. Dye, Y. Coulter, and R. E. Muenchausen, Appl. Phys. Lett. **62**, 2434 (1993).

[19] D.B. Kacedon, R.A. Rao, and C.B. Eom, Appl. Phys. Lett. **71**, 1724 (1997).

[20] G. Herranz, B. Martínez, J. Fontcuberta, F. Sánchez, C. Ferrater, M. V. García-Cuenca, and M. Varela, Phys. Rev. B **67**, 174423 (2003).

[21] A. Ohtomo and H.Y. Hwang, Nature **427**, 423 (2004).

[22] Q. Gan, R. A. Rao, C.B. Eom, L. Wu, and F. Tsui, J. Appl. Phys.**85**, 5297 (1999).

[23] M. Ziese, I. Vrejoiu, and D. Hesse, Phys. Rev. B **81**, 184418 (2010).

[24] See supplemental material at http://dx.doi.org/... for a detailed x-ray diffraction analysis and the temperature-dependent sheet resistance of all superlattice structures.

[25] C. B. Eom, R. J. Cava, R. M. Fleming, J. M. Phillips, R. B. van Dover, J. H. Marshall, J.W.P. Hsu, J.J. Krajewski, and W.F. Peck, Jr., Science **258**, 1766 (1992).

[26] P. Kostic, Y. Okada, N. C. Collins, Z. Schlesinger, J. Reiner, L. Klein, A. Kapitulnik, T. Geballe, and M. R. Beasley, Phys. Rev. Lett.**81**, 2498 (1998).

[27] G. Bergmann, Phys. Rev. B **28**, 515 (1983).

[28] Z. Q. Liu, D. P. Leusink, X. Wang, W. M. Lü, K. Gopinadhan, A. Annadi, Y. L. Zhao, X. H. Huang, S. W. Zeng, Z. Huang, A. Srivastava, S. Dhar, T. Venkatesan, and Ariando, Phys. Rev. Lett.**107**, 146802 (2011).

[29] L. A. Ponomarenko, A. K. Geim, A. A. Zhukov, R. Jalil, S.V. Morozov, K.S. Novoselov, I.V. Grigorieva, E.H. Hill, V.V. Cheianov, V.I. Fal'ko, K. Watanabe, T. Taniguchi, and R.V. Gorbachev, Nature Phys.**7**, 958 (2011).

[30] C. B. Bark, D. A. Felker, Y. Wang, Y. Zhang, H. W. Jang, C. M. Folkman, J. W. Park, S. H. Baek, H. Zhou, D. D. Fong, X. Q. Pan, E. Y. Tysmbal, M. S. Rzchnwski, and C. B. Eom, Proc. Natl. Acad. Sci. U.S.A. **108**, 4720 (2011).

[31] L. Klein, J. S. Dodge, C. H. Ahn, J. W. Reiner, L. Mieville, T. H. Geballe, M. R. Beasley, and A. Kapitulnik, J. Phys.: Condens. Matter **8**, 10111 (1996).

[32] R. Scherwitzl, S. Gariglio, M. Gabay, P. Zubko, M. Gibert, and J.-M. Triscone, Phys. Rev. Lett. **106**, 246403 (2011).

**Figure Captions:**

FIG. 1. (a) AFM image of a $TiO_2$-terminated STO single crystal substrate. (b) $\theta$-$2\theta$ scan spectroscopy of a SL7 superlattice grown on STO. (c) Reciprocal space mapping of the SL7 superlattice in the vicinity of the STO (-103) Bragg peak.

FIG. 2. $R_s$-$T$ curves of a 50 nm SRO single-layer film in (a) as well as SL7 in (b), SL4 in (c) and SL2 superlattices in (d) on different scales both under a zero-field and an out-of-plane 9 T magnetic field.

FIG. 3. (a) Temperature dependence of magnetization of different samples measured by an in-plane 1000 Oe magnetic field. (b) Out-of-plane magnetoresistance of a SL2 superlattice. (c) Field-dependent magnetic moment of the SL2 superlattice.

FIG. 4. Field effect transistors based on SRO/LAO superlattices. (a) Schematic of a field effect device based on SRO/LAO structure. Sheet resistance of the SL2 (b) and SL7 (c) superlattices as a function of back gate voltage.

**TABLE I** Characteristic resistance upturn temperatures and transport categories of SRO/LAO superlattices grown on different substrates

| Substrate | SL2 | SL3 | SL4 | SL7 |
|---|---|---|---|---|
| LSAT ($a$=3.868Å) | SL - >300 K | MIT - 196 K | MIT - 175 K | WL - 36 K |
| STO ($a$=3.905 Å) | SL - >300 K | MIT - 185 K | MIT - 185 K | WL - 50 K |
| DSC ($a$=3.944 Å) | - | SL - >300 K | MIT - 202 K | WL - 20 K |

Where SL, MIT and WL represent strong localization, metal-insulator transition and weak localization, respectively.

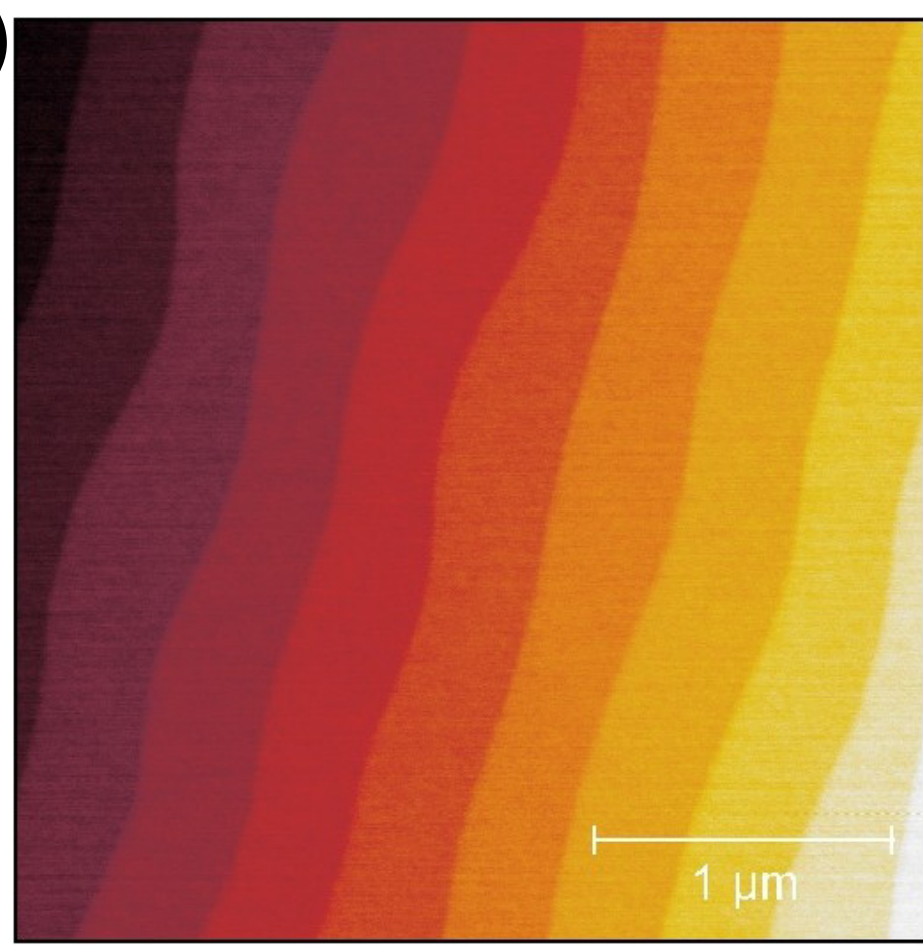 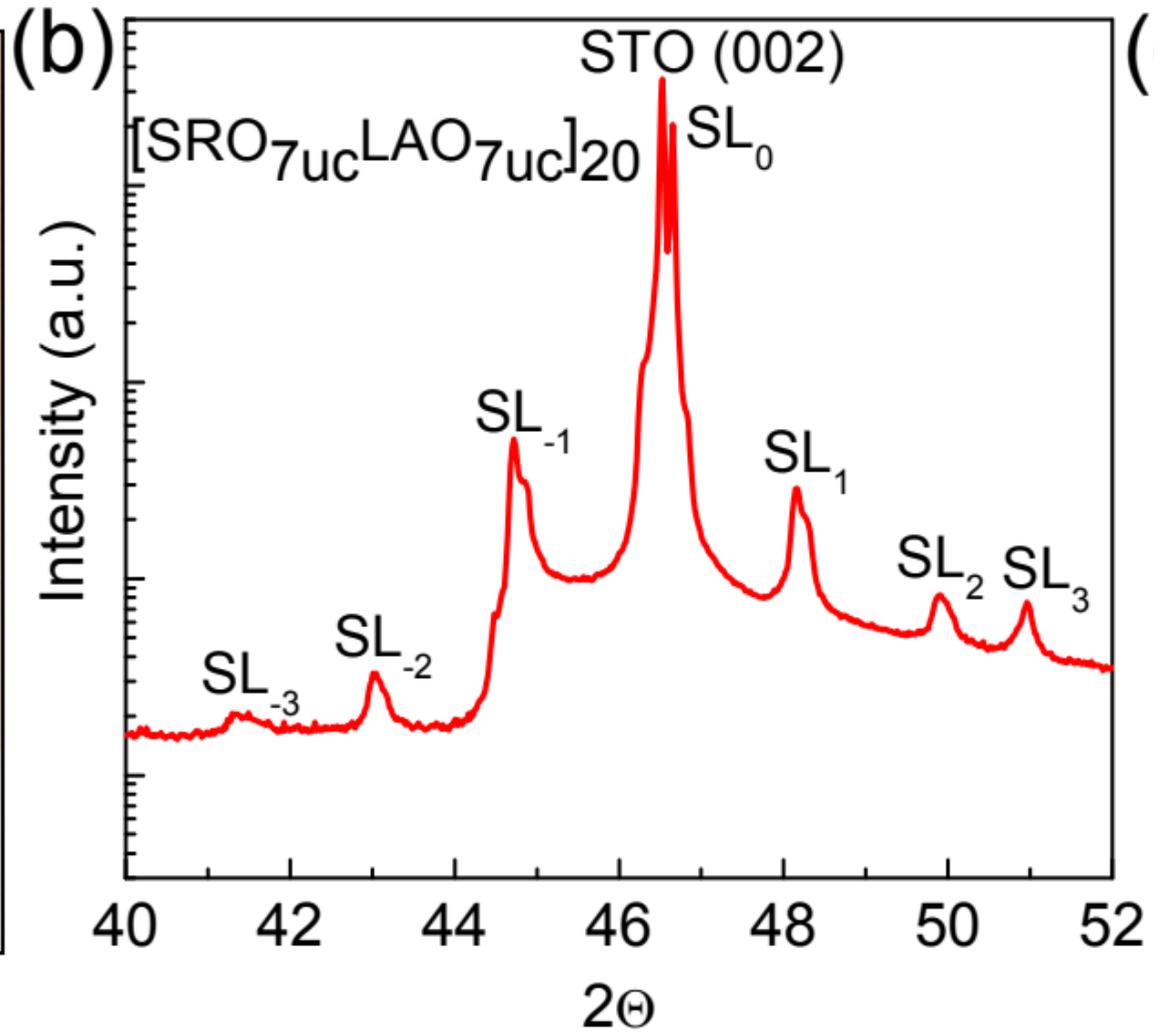 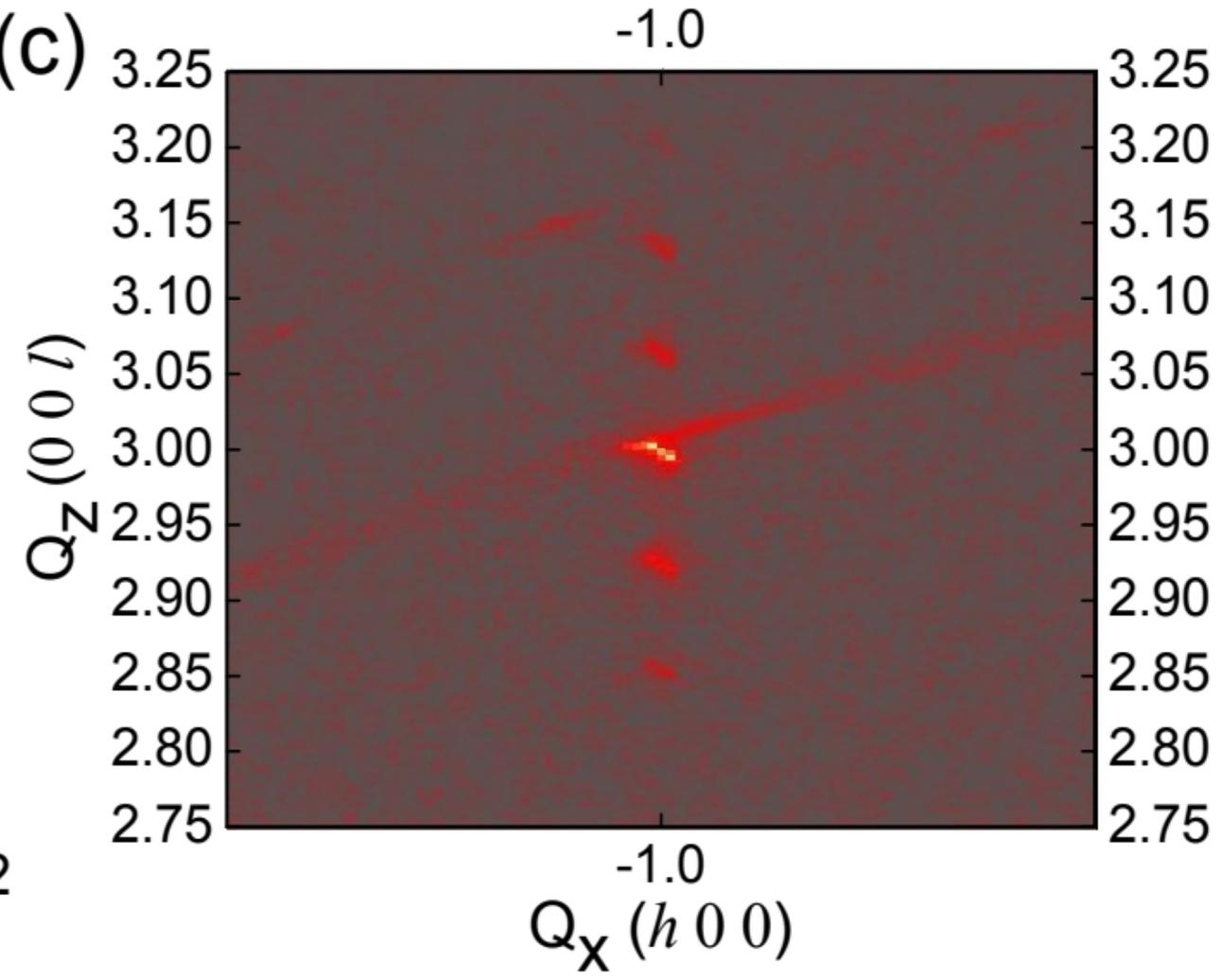

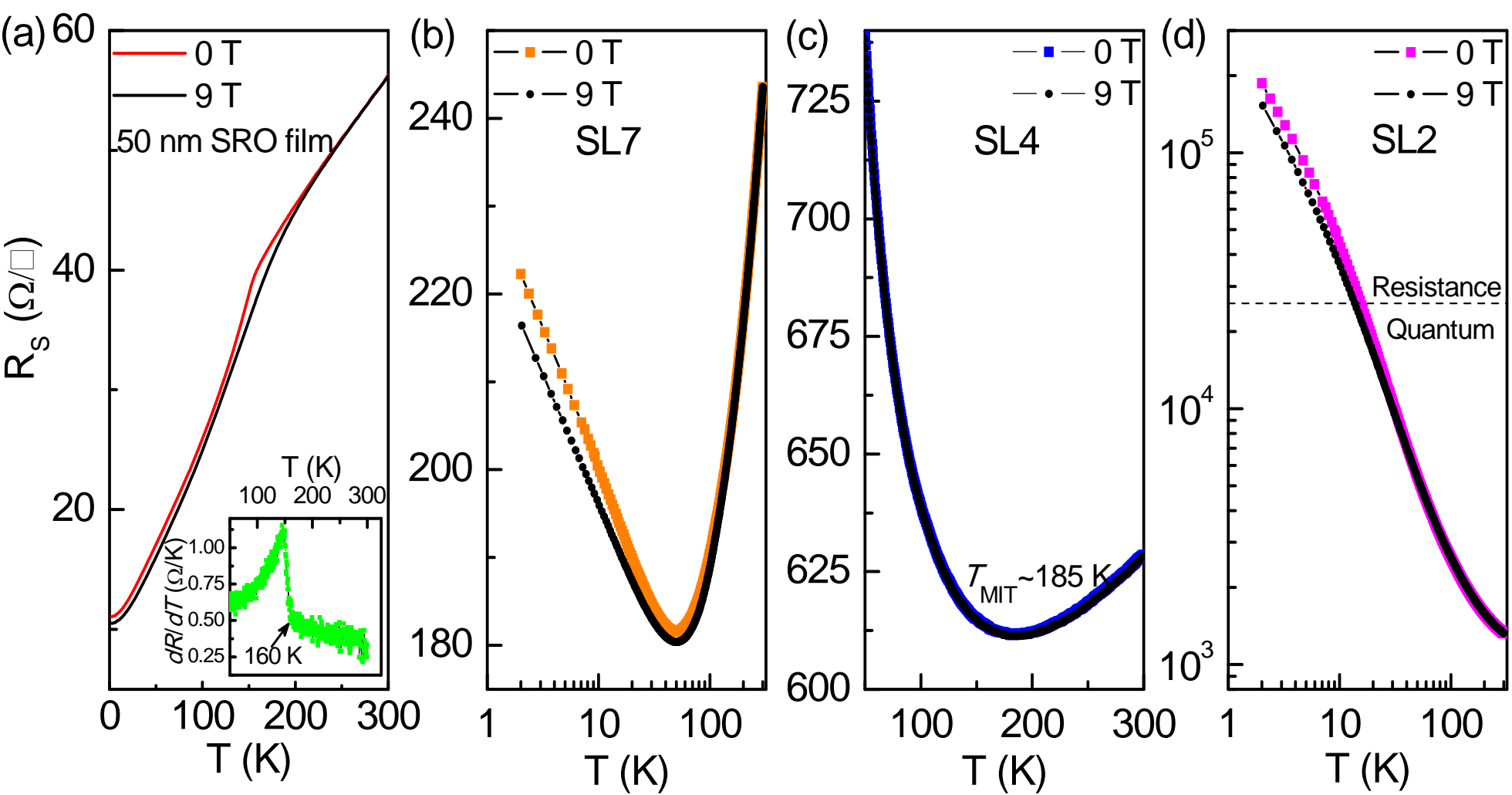

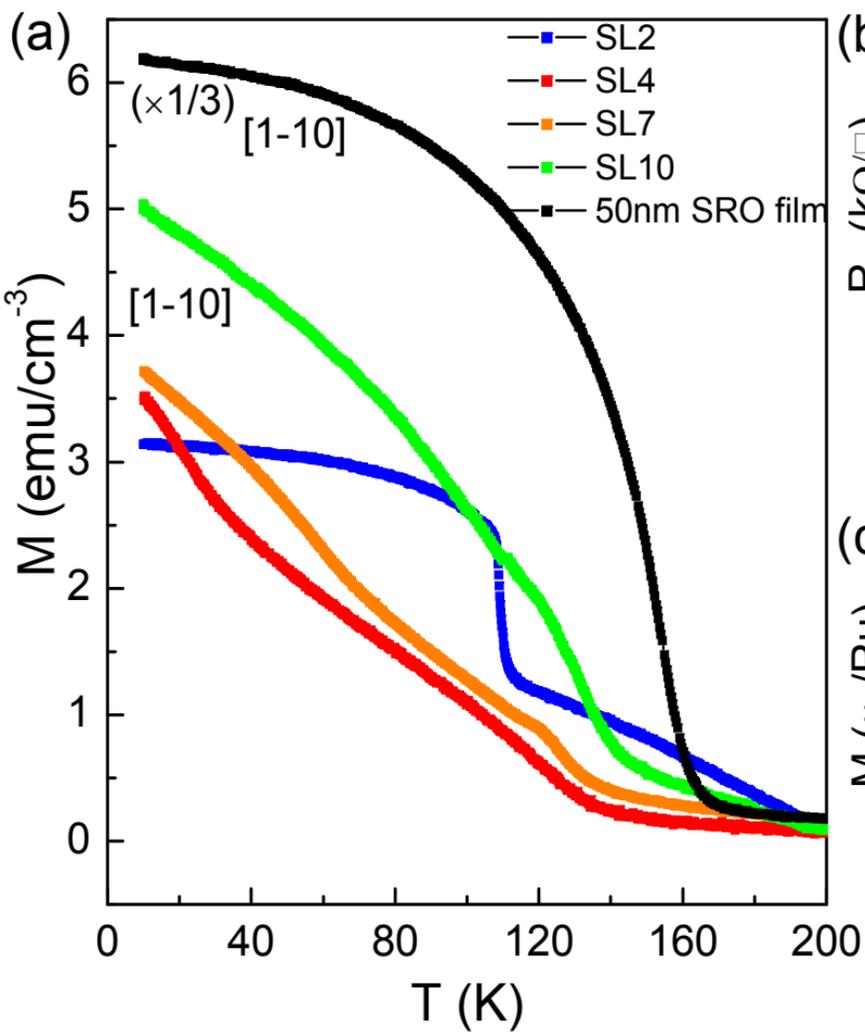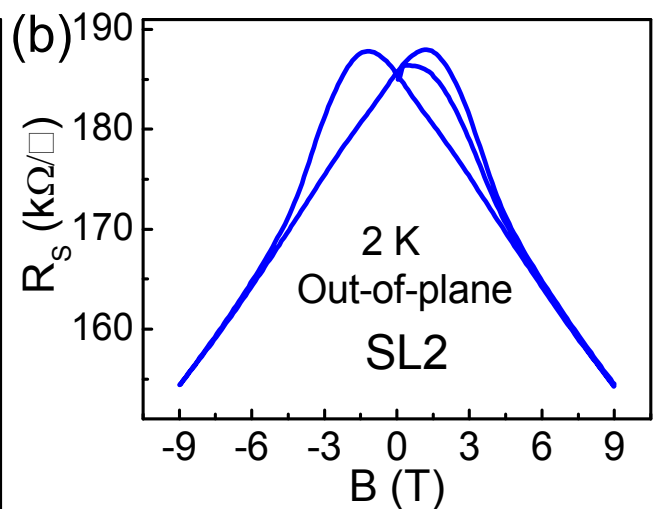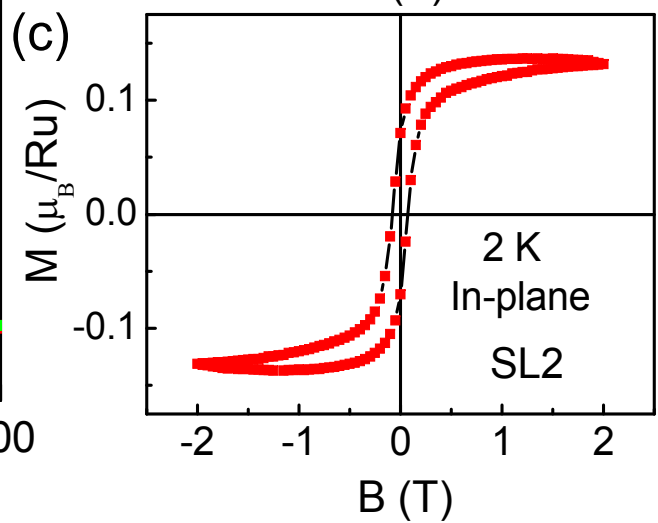

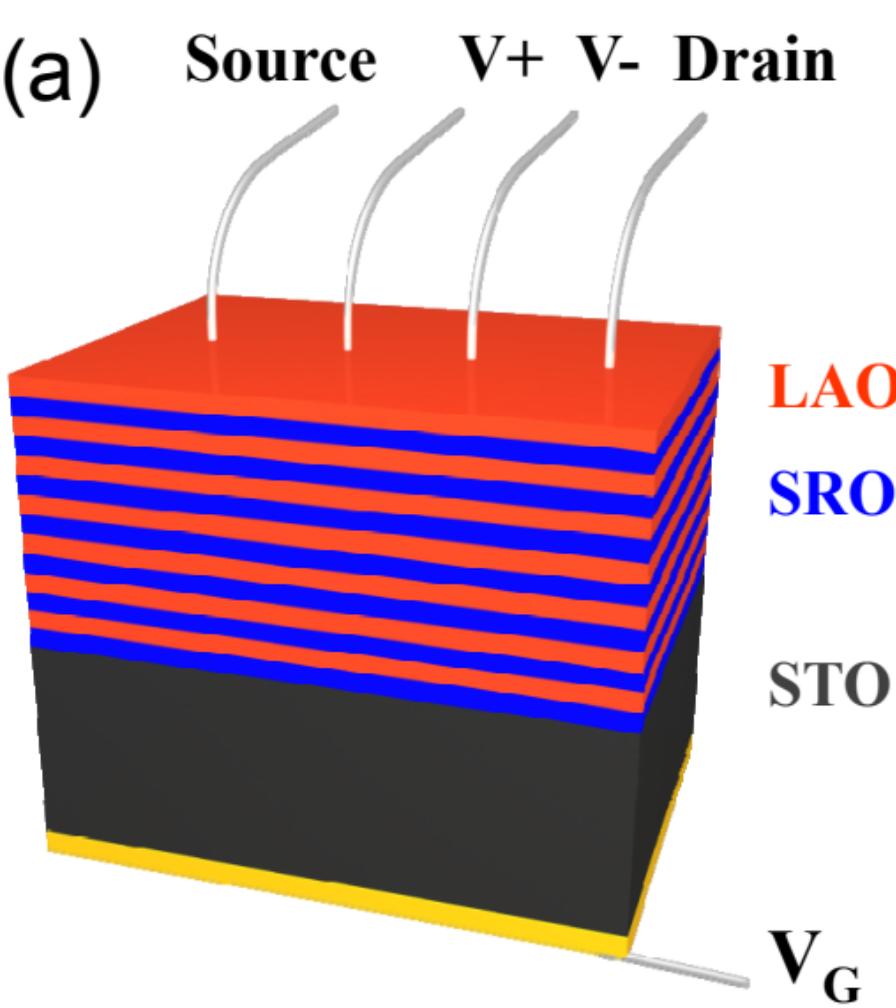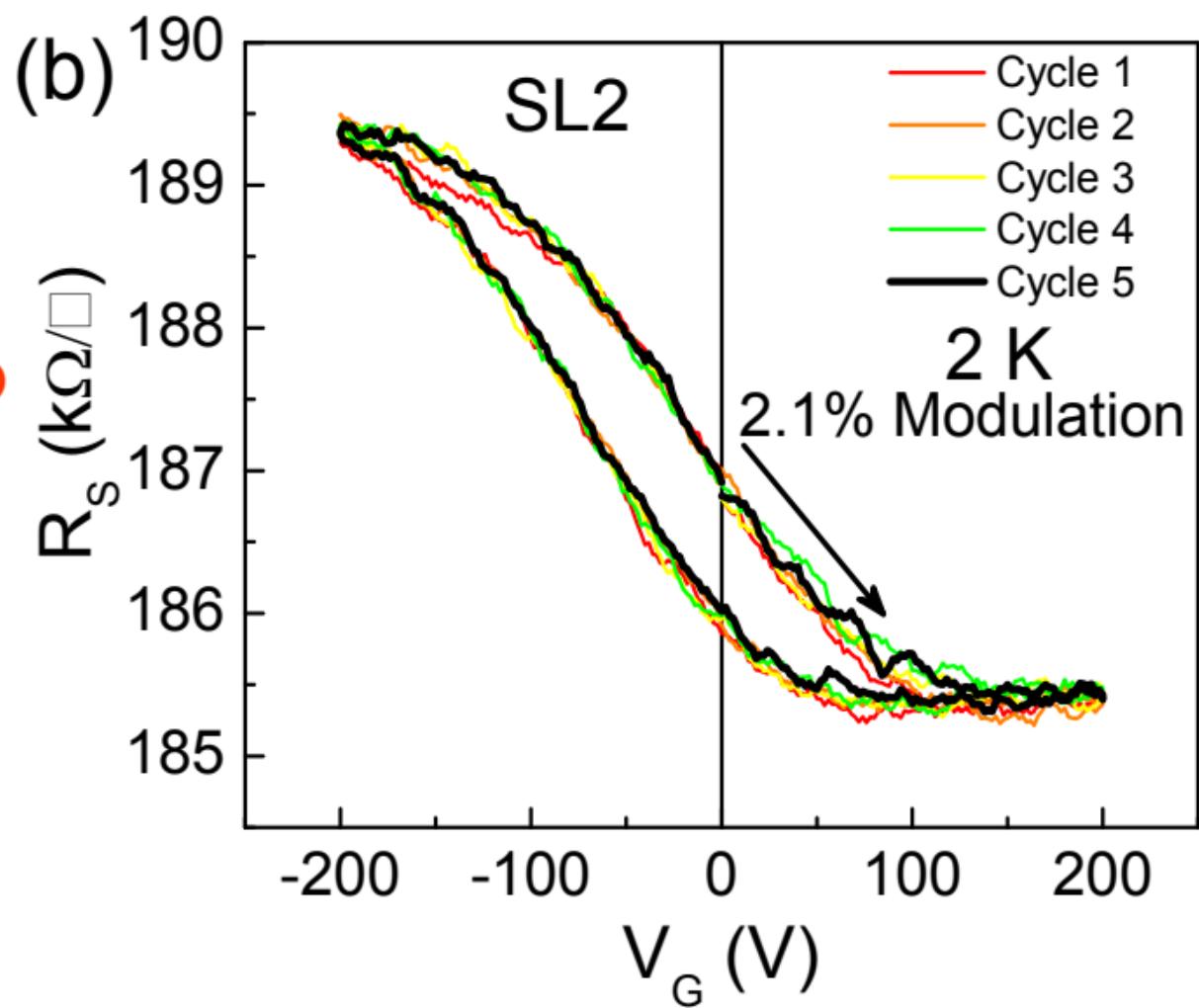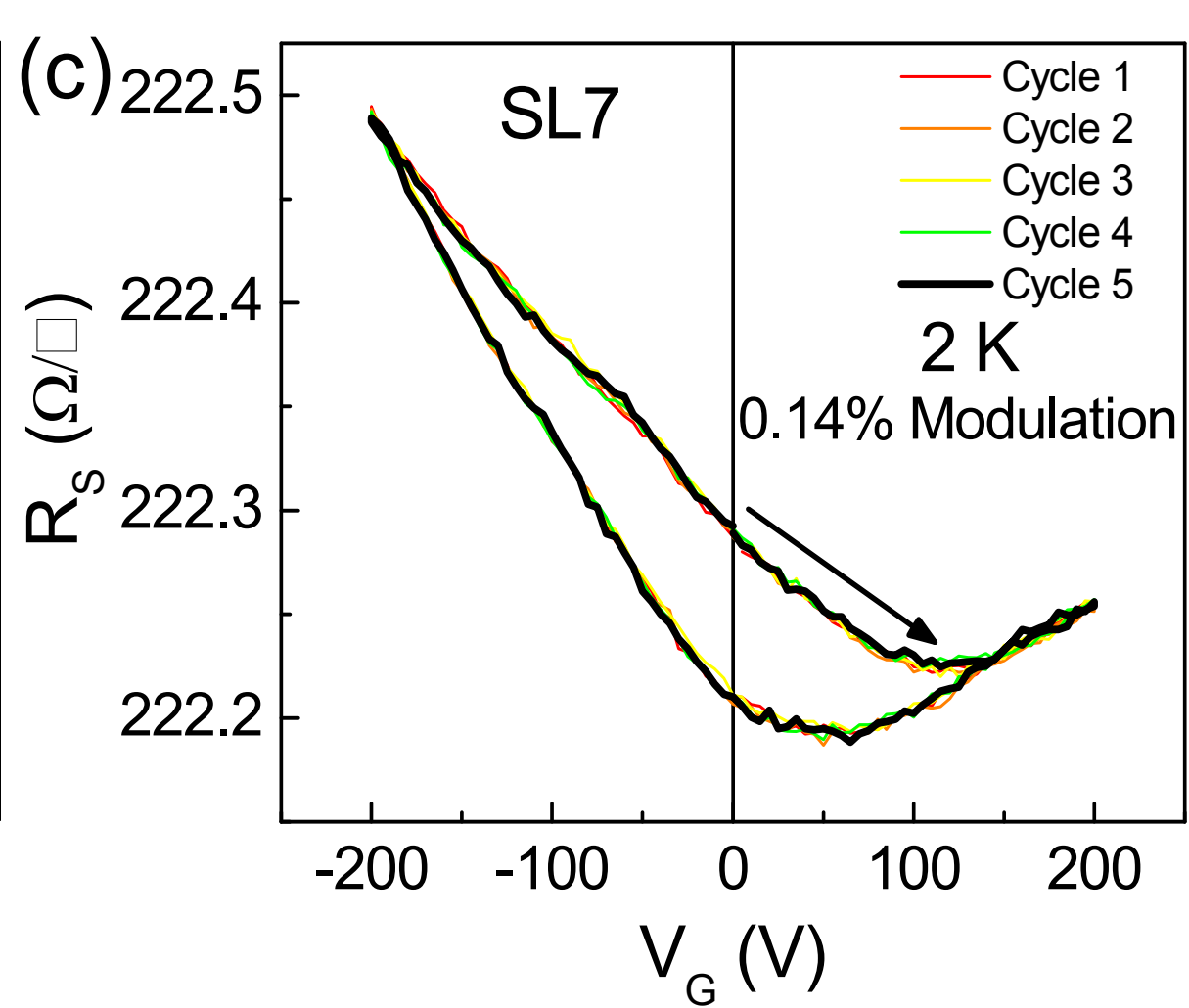

**Supplemental Material for "Tailoring the electronic properties of SrRuO$_3$ films in SrRuO$_3$/LaAlO$_3$ superlattices"**

Z. Q. Liu,[1,2] M. Yang,[2] W. M. Lü,[1,3] Z. Huang,[1] X. Wang,[1,2] B. M. Zhang,[1,4] C. J. Li,[1,5] K. Gopinadhan,[1,3] S. W. Zeng,[1,2] A. Annadi,[1,2] Y. P. Feng,[1] T. Venkatesan,[1,2,3] and Ariando[1,2*]

[1]*NUSNNI-Nanocore, National University of Singapore, Singapore 117411*

[2]*Department of Physics, National University of Singapore, Singapore 117542*

[3]*Department of Electrical and Computer Engineering, National University of Singapore, Singapore 117542*

[4]*Departement of Material Science Engineering, National University of Singapore, Singapore 117576*

[5]*National University of Singapore (NUS) Graduate School for Integrative Sciences and Engineering, 28 Medical Drive, Singapore 117456.*

### 1. Detailed x-ray diffraction analysis on the [SRO$_{7uc}$LAO$_{7uc}$]$_{20}$ superlattice

As shown in Fig. 1(b) in the manuscript, the zero-order satellite peak of the superlattice is close to the STO (002) peak, which also confirms the studies by Gan *et al.*[1] and Ziese *et al.*[2] that SRO layers are (110)-oriented since the zero-order satellite peak corresponds to the average of the out-of-plane lattice constant. Furthermore, the thickness $\Lambda$ of each period and the average lattice constant $d$ of the superlattice can be fitted by satellite peak positions using the following equation:[3]

$$\frac{2\sin\theta}{\lambda} = \frac{1}{d} + \frac{n}{\Lambda}$$

where $\lambda$ is the wavelength of x-ray and $n$ is an integer representing the order of satellite peaks. A highly linear dependence of $n$ on $2\sin\theta/\lambda$ results in $\Lambda$ = 55.5 Å, which is comparable to the thickness of 7 uc SRO plus 7 uc LAO (3.93×7+3.79×7 Å=54 Å). The reciprocal space mapping of the SL7superlatticein the vicinity of the STO (-103) Bragg peak is shown in Fig. 1(c). Satellite peaks up to the third order are visible, which are consistent with the $\theta$-$2\theta$ scan spectroscopy. All the satellite peaks have the same projection onto the $Q_x$ axis as that of the STO Bragg peak indicating coherent epitaxial growth.

## 2. Sheet resistance of different superlattices as a function of temperature

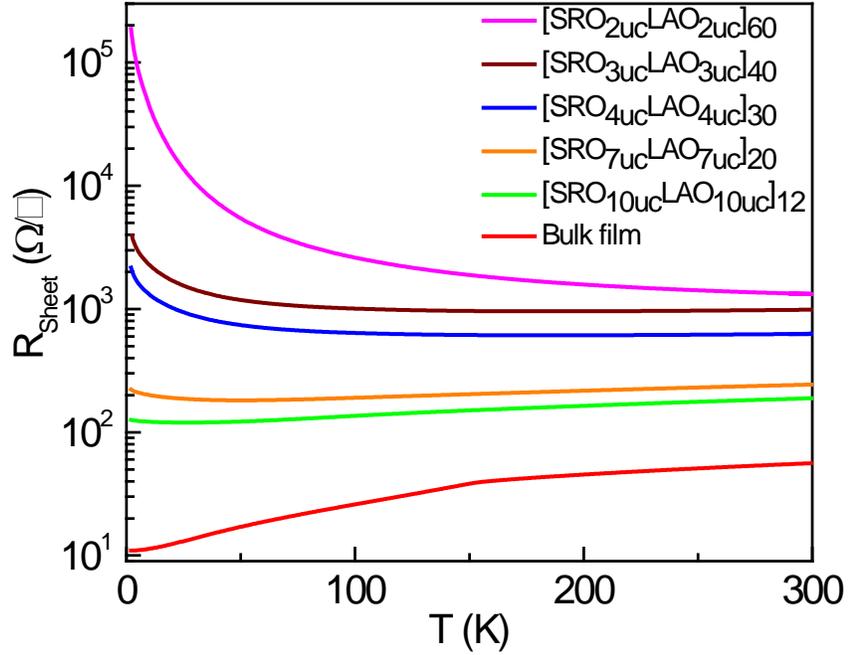

**Figure S1:** Temperature dependence of sheet resistance of different $SrRuO_3/LaAlO_3$ superlattice structures as well as a 50 nm $SrRuO_3$ film.

[1] Q. Gan, R. A. Rao, C.B. Eom, L. Wu, and F. Tsui, J. Appl. Phys. **85**, 5297 (1999).

[2] M. Ziese, I. Vrejoiu, and D. Hesse, Phys. Rev. B **81**, 184418 (2010).

[3] E. E. Fullerton, I. K. Schuller, H. Vanderstraeten, and Y. Bruynseraede, Phys. Rev. B 45, 9292 (1992).